\documentclass[12pt]{article}
\usepackage[body={17.5cm, 22cm},right=2cm]{geometry}
\usepackage{amssymb,graphicx}
\usepackage{amsmath}
\usepackage{epsfig}

\newcommand{\be}{\begin{equation}}
\newcommand{\ee}{\end{equation}}
\newcommand{\bea}{\begin{eqnarray}}
\newcommand{\eea}{\end{eqnarray}}
\newcommand{\bg}{\begin{gather}}
\newcommand{\eg}{\end{gather}}
\newcommand{\bseq}{\begin{subequations}}
\newcommand{\eseq}{\end{subequations}}

\newcommand{\loe}{\lesssim}

\def\slash#1{\setbox0=\hbox{$#1$}#1\hskip-\wd0\hbox to\wd0{\hss\sl/\/\hss}}

\begin{document}
\begin{titlepage}
\begin{flushright} OUTP\,-\,09/23P \end{flushright}
\begin{center}
{\LARGE\bf  
String Photini at the LHC\\}
\vspace{0.5cm}
{ \large
Asimina Arvanitaki$^{a,b}$,
Nathaniel Craig$^c$,
Savas Dimopoulos$^c$, \\ 
\vspace{.3cm} Sergei~Dubovsky$^{c,d}$, and John March-Russell$^e$}\\
\vspace{.45cm}
{\small  \textit{  $^{\rm a}$ Berkeley Center for Theoretical Physics, University of California, Berkeley, CA, 94720}}\\ 
\vspace{.1cm}
{\small  \textit{  $^{\rm b}$ Theoretical Physics Group, Lawrence Berkeley National Laboratory, Berkeley, CA, 94720 }}\\ 
\vspace{.1cm}
{\small  \textit{  $^{\rm c}$ Department of Physics, Stanford University, Stanford, CA 94305, USA }}\\ 
\vspace{.1cm}
{\small  \textit{  $^{\rm d}$
Institute for Nuclear Research of the Russian Academy of Sciences, 
        60th October Anniversary Prospect, 7a, 117312 Moscow, Russia}}\\
\vspace{.1cm}
{\small  \textit{  $^{\rm e}$
Rudolf Peierls Centre for Theoretical Physics, University of Oxford, 1 Keble Road, Oxford, UK
}}
\end{center}
\begin{center}
\begin{abstract}
String theories with topologically complex compactification manifolds suggest the simultaneous presence of many unbroken U(1)'s without any light matter charged under them. The gauge bosons associated with these U(1)'s do not have  direct observational consequences. However, in the presence of low energy supersymmetry the gauge fermions associated with these U(1)'s, the ``photini", mix with the Bino and extend the MSSM neutralino sector. This leads to novel signatures at the LHC. The lightest ordinary supersymmetric particle (LOSP) can decay to any one of these photini. In turn, photini may  transition into each other, leading to high lepton and jet multiplicities. Both the LOSP decays and inter-photini transitions can lead to displaced vertices. When the interphotini decays happen outside the detector, the cascades can result in different photini escaping the detector leading to multiple reconstructed masses for the invisible particle. If the LOSP is charged, it stops in the detector and decays out-of-time to photini, with the possibility that the produced final photini vary from event to event.  Observation of a plenitude of photini at the LHC would be evidence that we live in a string vacuum with a topologically rich compactification manifold.
\end{abstract}
\end{center}
\end{titlepage}

\section{Introduction and Summary}
String theory is a mathematically successful and beautiful theory of quantum gravity. However, as is natural to expect for any theory of quantum gravity given the enormous value  of the relevant energy scale, $M_{Pl}\simeq 10^{19}$~GeV, testing string theory at experimentally accessible energies is challenging. Two major qualitative predictions of string theory are supersymmetry (SUSY) and extra spatial dimensions. The weak hierarchy problem  suggests that at least one of these phenomena may be accessible to observations at TeV energies.

The discovery of large extra dimensions at the LHC would certainly open a spectacular window into string dynamics. Here we concentrate on a more challenging scenario in which the weak hierarchy problem is solved by low energy SUSY, but the size of extra dimensions is very small (for instance, Planck or GUT scale). Although the very discovery of low energy SUSY would provide strong support for the string framework, it is natural in this case to ask whether further evidence for string theory at low energies may exist.

The usual characteristic signature of extra dimensions---excited Kaluza--Klein (KK) modes---is unavailable for small extra dimensions, as massive KK modes are too heavy to be produced. However, realistic string theory constructions typically result in extra-dimensional manifolds with rich and non-trivial topology.  One way to characterize the topological complexity of the manifold is by enumerating closed sub-manifolds(cycles) of different dimensionality that cannot be deformed one into another---the so-called (co)homology classes. This is a natural generalization of the way in which orientable closed two-dimensional surfaces can be characterized by the number of handles.  Realistic compactifications in string theory typically involve manifolds with a large number of cycles---from several hundreds to $10^5$. The reason for this is simple combinatorics---generically there are many non-equivalent ways to embed a lower dimensional surface in a reasonably non-trivial six-dimensional manifold.  For instance, the simplest Calabi--Yau space---a six torus---has six 1- and 5-cycles, fifteen 2- and 4-cycles, and twenty 3-cycles.

Interestingly, the topological complexity of a compactification manifold leaves imprints in the spectrum of KK zero modes, even if the size of the extra dimensions is arbitrarily small. To understand how this happens, let us recall another intrinsic feature of string theory: the presence of a new kind of gauge field, in the guise of antisymmetric tensor fields (forms) of various rank. In four space-time dimensions an antisymmetric second rank tensor (2-form) $B_{\mu\nu}$ is equivalent to a massless scalar field, while higher rank forms are non-dynamical. This changes in higher dimensions, where higher rank antisymmetric tensor fields can be both dynamical and different from the scalar (0-form), and vector
(1-form) fields. Higher rank antisymmetric forms play a crucial role in the Green--Schwarz mechanism for anomaly cancellation in string theory and are related to the presence of extended objects in the theory such as strings and branes. Just as a vector field is coupled to the world-line of a charged particle, higher rank forms are coupled to the world-volumes of extended objects. Of particular interest in what follows are the Ramond--Ramond (RR) forms $C_{2,4}$ of type IIB theory of rank 2 and 4, or $C_{1,3}$ of type IIA theory of rank 1 and 3; the extended objects charged under these fields are D-branes \cite{Polchinski:1995mt}.

The crucial property of antisymmetric tensor fields is that upon compactification they give rise to {\it many} KK zero modes, labeled by the independent cycles of the internal manifold \cite{Green:1987mn}.  Interestingly, the number of zero modes depends only on the topology of extra dimensions, but not on their absolute size.  Indeed, zero modes are scale free, so that their number cannot depend on a dimensionful parameter.  Consequently, zero modes provide a probe of extra dimensions even in the limit where their size is tiny. The discovery of a large number of particles with similar properties whose presence is hard to motivate within a strictly 4-dimensional theory would be evidence for the existence of extra dimensions with complicated topology\footnote{Replication of the Standard Model generations may already be a hint supporting this logic.}.
For instance, every independent 4-cycle $\Sigma^4_i$ in type IIB string theory gives rise to an ultra-light\footnote{Being massless at the perturbative level, these fields acquire a mass due only to non-perturbative effects.} (pseudo)scalar field with axion-like couplings, defined as an integral of $C_4$ over the 4-cycle.
More generally, every independent $n$-cycle gives rise to a scalar KK zero mode in the presence of a rank $n$ form.

One of these pseudoscalar fields may play the role of the QCD axion \cite{Svrcek:2006yi}, while others may be observed by a number of cosmological and astrophysical experiments in the next decade \cite{axiverse}. It is worth keeping in mind that these string axions may acquire a high-scale mass in a number of ways (e.g., due to the presence of branes wrapping the corresponding cycles or from fluxes; they may also be projected away by orientifold planes). However, the strong CP problem suggests that at least one of these fields survives at low energies. Given the large number of independent cycles on a typical compactification manifold, it would be strange if only one of them gave rise to the light axion, thereby leading to the expectation of a plenitude of ultra-light axion-like particles--- the ``string axiverse''   \cite{axiverse}. 

String axions are (pseudo)Goldstone bosons and cannot have any renormalizable couplings with the fields of the Standard Model. All their interactions are suppressed by the compactification scale, so there is no opportunity to observe string axions in conventional collider experiments. However, these string axions are not the only matter suggested by a topologically-complex compactification manifold.  The main point of the current paper is that in the string axiverse with low energy SUSY it is natural to expect another plenitude of particles with weak scale masses that {\it can} be observed at the LHC.

The reason is that an antisymmetric form of rank $n$ gives rise also to massless vector fields, labeled by the independent cycles of dimension $(n-1)$. As in the scalar case, these vectors are defined as integrals of the form over the corresponding cycle. For instance, in type IIB theory each of the 3-cycles $\Sigma^3_i$ makes it possible to define a 4d vector field
\be
A_\mu^i = \int_{\Sigma^3_i} C_4
\label{U1def}
\ee
by taking three of the four-form indices along the directions of the cycle. Moreover, each 4d vector
field $A_\mu^i$ inherits a gauge symmetry from the underlying 10d abelian gauge symmetry of the RR
field $C_4\to C_4 +d \Lambda_3$, so that the end result is a plethora of 4d $U(1)$ gauge fields. 

As with string axions, these fields may acquire a high mass from fluxes, or may be projected away by orientifold planes.  However, as before, there is no reason for this to occur with all such vector fields and it is therefore natural to expect a plenitude of massless $U(1)$ fields in the string axiverse.  It is interesting to note that essentially the same ingredients---cycles and form fields---give rise to the string landscape of vacua that motivates fine-tuning of the vacuum energy.  Successful scanning of the vacuum energy suggests the presence of at least several hundreds of cycles (giving rise to the famous $10^{{\rm few}\times100}$ vacua of string theory), thus providing an additional motivation for the plenitude of axions/photons.

An important property of the string RR $U(1)$ fields is that there are generically no light states charged under them. The reason is that the only objects charged under RR forms are non-perturbative D-brane states, so that particle states charged under RR $U(1)$'s arise from D-branes wrapping the corresponding cycles. These states  have masses above the string scale apart from the exceptional case of vanishingly small cycle volumes.

As a result, at low energies the RR $U(1)$'s interact with the Standard Model fields---which themselves arise from light perturbative string states---either through higher-dimensional operators unobservable at  colliders, or through renormalizable kinetic mixing terms with the hypercharge $U(1)_Y$\footnote{As observational consequences
are the main focus of this paper, we postpone the discussion of the origin of the mixing in string theory in the RR case until section \ref{discussion}.}.  In the presence of light particles charged under additional $U(1)$'s such kinetic mixing would be strongly constrained from astrophysics and laboratory searches for millicharged particles \cite{Davidson:1993sj, Davidson:2000hf,Dubovsky:2003yn,Melchiorri:2007sq}.
However, as discussed above, such light millicharged states are absent for RR $U(1)$'s and these constraints do not pertain. As a consequence of the absence of light charged states, kinetic mixings with RR photons can be removed by field redefinition in the low energy theory without introducing any physical effects apart from changing the value of the hypercharge gauge coupling.  Consequently, massless RR vector fields {\it per se} do not provide a useful observational window into extra dimensions.

However, the situation becomes significantly more interesting in the presence of low energy SUSY.
In this case massless RR photons are accompanied by their light fermionic superpartners---photini.
Unlike vectors, RR photini acquire masses of order the gravitino mass $m_{3/2}$ as a result of SUSY breaking.  If the dominant source of SUSY breaking for the MSSM also comes from the gravity mediation, then these photini masses are of the same order as the MSSM soft masses.  This is the most interesting case for the LHC, and therefore will remain our primary focus in what follows.  Another possibility is that the dominant source for the communication of SUSY breaking to the MSSM comes from gauge mediation, so that RR photini are much lighter than the MSSM superpartners.

As a consequence of a non-trivial photini mass matrix, the mixing of RR photini with the bino cannot be rotated away and has observable effects as we discuss in Section \ref{pheno}.   For the purposes of LHC phenomenology, the significant result of this mixing is the extension of the MSSM neutralino sector by a plenitude of new states mixed with the bino through the gaugino mass matrix.  This leads to a variety of possible signatures depending on the amount of mixing and the size of the inter-photini mass splittings, including extended supersymmetric cascades with high lepton and jet multiplicities arising from inter-photini transitions; displaced vertices from Lightest Ordinary Supersymmetric Particle (LOSP) decays or inter-photini transitions; cascades ending with different photini escaping the detector leading to multiple reconstructed masses for the invisible particle; and if the LOSP is charged so it stops, out-of-time decays of the LOSP to photini, with the possibility of the produced photini varying from event to event.   Combinations of these signatures can also coexist.  

Finally we emphasize that these photini signatures can occur for any set of $U(1)$'s that kinetically mix with hypercharge and do not possess light charged states, not just the photini of RR $U(1)$'s \cite{Ibarra:2008kn}.  Such multiple hidden $U(1)$'s are not uncommon in string theory and can arise from a variety of sources--for example, isolated branes not intersecting with the branes that realize the SM sector.  If the isolated brane possess only vector-like matter---the more common case---the matter can get a large positive mass--squared leaving an unbroken $U(1)$ with no surviving light charged states.  

Of course, the existence of a plenitude of (possibly very) weakly coupled photini may pose challenges for conventional cosmology. In Section \ref{cosmo} we consider the potential constraints on photini masses and mixings from cosmological considerations, as well as the various means by which these constraints may be obviated. In Section \ref{totalitarian} we turn to the case of light photini in theories with gauge mediated SUSY breaking. Although the prospective signatures of such light photini at the LHC are less promising, their decays may give rise to observable astrophysical signals. 


\section{Phenomenology}\label{pheno}
\subsection{The photino lagrangian}
Let us now turn to the 4d effective theory arising from kinetic mixing between visible and hidden gauge sectors. It has been well known for many years \cite{Holdom:1985ag} that theories with multiple $U(1)$ gauge symmetries may admit kinetic mixings among the different $U(1)$'s. Consider, for simplicity, the case of two such symmetries, $U(1)_a \times U(1)_b$. For the typical case of interest, $U(1)_a$ is a visible-sector gauge symmetry such as hypercharge $U(1)_Y$, while $U(1)_b$ is some hidden-sector gauge symmetry. In the basis in which  the interaction terms have the canonical form, the pure gauge part of the Lagrangian can be written as
\be
\mathcal{L}_{\rm gauge}~=~ -\frac{1}{4}\,F_{(a)}^{\mu\nu}F_{(a)\mu\nu}
-\frac{1}{4}\,F_{(b)}^{\mu\nu}F_{(b)\mu\nu}
+\frac{\epsilon}{2}\,F_{(a)}^{\mu\nu}F_{(b)\mu\nu}~.
\label{firstequation}
\ee
where $\epsilon$ parametrizes the kinetic mixing between the two $U(1)$s. In a supersymmetric theory, such a Lagrangian generalizes to \cite{Dienes:1996zr}
\be
  \mathcal{L}_{\rm gauge}~=~ \frac{1}{32}\int d^2\theta \,\left\{W_aW_a+W_bW_b-2\epsilon W_aW_b
\right\}
\ee
where $W_a$ and $W_b$ are the chiral gauge field strength superfields for
the two gauge symmetries (e.g., $W_a =\bar D^2D V_a$ for the $U(1)_a$ vector superfield $V_a$). To bring the pure gauge portion of the Lagrangian to canonical form, we may shift the hidden-sector gauge field via
\be
V^\mu_b~\to~ V'^\mu_b=V^\mu_b - \epsilon V^\mu_a
\ee
so that $W_b \to W'_b=W_b- \epsilon W_a$. 
This renders the gauge Lagrangian diagonal,
\be
\mathcal{L}_{\rm gauge}~=~\frac{1}{32}\int d^2\theta \,\left\{W_a W_a+ W_b' W_b' \right\}~,
\ee
and shifts the visible-sector gauge coupling by an amount
\be
g_a \rightarrow g_a / \sqrt{1 - \epsilon^2}~.
\label{couplingshift}
\ee

If there are no light states charged only under $U(1)_b,$ then the above field redefinition produces no change in the interactions of states charged only under $U(1)_a.$  Thus the theory is relatively uninteresting in the absence of light hidden-sector charged states; the hidden sector photon decouples entirely, the only remnant being
the shift in the hypercharge gauge coupling \cite{Holdom:1990xp}.  The success of supersymmetric gauge coupling unification, if assumed to be non-accidental, then indicates that $\sum_{i}\epsilon_i^2 \loe 0.01$, where the sum runs over
all $U(1)$'s with which hypercharge mixes.       

However, the hidden sector gaugino $\lambda_b$ may not decouple so readily when supersymmetry is broken. Although the $U(1)_b$ gauge boson may be decoupled by field redefinitions, the gaugino $\lambda_b$ still mixes with the visible sector via off-diagonal terms in the gaugino mass matrix. These remnant interactions between hidden-sector gauginos and visible-sector states provide indications of the hidden-sector gauge symmetry even in the absence of light states charged directly under $U(1)_b$.

Motivated by the appearance of many hidden-sector $U(1)$s arising from dimensional reduction of RR forms, let us now consider $n$ hidden-sector $U(1)$s kinetically mixed with the Standard Model hypercharge $U(1)_Y.$ The gauge bosons $A_\mu^i$ mix among themselves and with the hypercharge gauge boson $B_\mu$ via kinetic mixing, while the photini $\tilde \gamma_i$ mix among themselves and with the bino $\tilde B$ via both kinetic mixing and off-diagonal terms in the gaugino mass matrix. The structure of this mixing is determined by (among other things) the details of supersymmetry breaking, the geometry of the internal manifold, and induced mixing between brane and bulk gauge supermultiplets. 

The gauge kinetic terms may be rendered canonical by hidden-sector field redefinitions analogous to those discussed above. In the absence of light charged states, the canonically normalized $U(1)$ gauge fields $A_{\mu}^i$ and their $D$-terms decouple entirely. The only remnant impact on the hypercharge gauge boson $B_\mu$ is a shift in the hypercharge gauge coupling, which may have implications for unification when the mixings are large. 

The interesting physics lies in the photini. Despite the decoupling of the hidden-sector $U(1)$ gauge bosons, we crucially retain mixing in the gaugino mass matrix. The mixings between the photini and MSSM gauginos are encoded in the Lagrangian terms 
\be
\delta \mathcal{L} \supset i Z_{IJ} \lambda_I^\dag \slash \partial \lambda_J + m_{IJ}  \lambda_I  \lambda_J
\ee
where here $I,J$ run across the bino $\tilde B$ and $n$ photini $\tilde \gamma_i;$ the $Z_{IJ}$ encode arbitrary kinetic mixing, while the $m_{IJ}$ are generated when supersymmetry is broken. As with the gauge kinetic terms, the gaugino kinetic terms may be diagonalized via field redefinitions so that $Z_{IJ} \rightarrow \delta_{IJ}$ and $m_{IJ} \rightarrow m_{IJ}'$. In particular, if the kinetic terms can be made canonical by the transformation $\lambda_I \rightarrow \lambda'_I = P_{IJ}^{-1} \lambda_J,$ then $m_{IJ}' = P^{\dag}_{IK} m_{KL} P_{LJ}.$ It bears mentioning that if the original mass mixing terms are strictly proportional to the gauge kinetic term, then the mass mixing in the canonical basis vanishes. The persistence of mixing among gauginos requires that SUSY-breaking gaugino masses are not exactly proportional to the gauge kinetic mixing matrix, which has implications for the precise mechanism by which supersymmetry is broken and communicated to the gauginos.
Moreover since the final physical mixing among the gauginos depends on the mass
matrix mixing, the gauge-coupling unification constraint on the amount of kinetic mixing with hypercharge does not limit the size of the mixing among gauginos.

To study the neutralino mass eigenstates, we may diagonalize the gaugino mass matrix $\mathbf{m}$ via $\mathbf{m}_{D} = \mathbf{f}^* \mathbf{m} \, \mathbf{f}^{-1},$ where $\mathbf{f}$ is a unitary matrix. The mass eigenstate neutralinos $\tilde N_I$ may then be written as 
\be
\tilde N_I = f_{IJ} \lambda_J
\ee
where $I, J = 1, ..., n+4$ runs over the four MSSM neutralinos and the $n$ photini; $f_{IJ}$ are the components of the matrix $\mathbf{f},$ and $\lambda_I = (\tilde B, \tilde W, \tilde H_d, \tilde H_u, \tilde \gamma_1, ..., \tilde \gamma_n)$ are the gauge eigenstate gauginos with canonical kinetic terms.

When mixings are large, there is no particular distinction among neutralinos; every neutralino mass eigenstate is an admixture of MSSM gauginos and photini. In the limit of small mixing, however, the neutralinos decompose into mostly-MSSM and mostly-photino states. Consequently, we may think of the $\tilde N_a \, (a = 1, ..., 4)$ as mostly-MSSM neutralinos, and the $\tilde N_i \, (i = 5, ...., n+4)$ as mostly-photino neutralinos. For the sake of clarity we will henceforth concern ourselves primarily with the case of small mixings, though in principle a broad range of hidden-visible mixings may arise.

In the limit of small mixing, the components in $\mathbf{f}$ decompose accordingly: the coefficients $f_{ab}$ are akin to those of the conventional MSSM neutralino matrix and depend principally on the parameters $m_Z, \tan \beta, \mu, m_1, m_2.$ The coefficients $f_{i1},$ in turn, encode mixing between the hidden-sector photini and the bino. For simplicity, we will henceforth write $f_{i1} \equiv \epsilon_i.$ It is this mixing that gives rise to interactions between hidden-sector photini and the fields of the MSSM. It is important to emphasize that these $\epsilon_i$ are not identical to the original kinetic mixing terms $\epsilon_i W_i W_Y;$ they incorporate additional $\mathcal{O}(m_i/m_B)$ factors from the diagonalization of kinetic terms and the gaugino mass matrix.  

In the limit of small $\epsilon_i,$ the mixings between photini and the higgsinos (and wino) are of order $f_{i (2,3,4)} \simeq  f_{1(2,3,4)} \epsilon_i,$ and may be parametrically smaller than the photino-wino mixing by MSSM mixings. Lastly, the coefficients $f_{ij}$ correspond to mixings among the various photini, and vary from  $\sim 10^{-3} - 1$ depending on the range of cycle areas and their intersection properties.\footnote{The mixings among photini are dictated by the gauge kinetic coupling matrix for the RR fields; at tree level this takes the form $\mathbf{Z}_{RR} \propto \mathbf{A} \mathbf{C}^{-1} + i \mathbf{C}^{-1},$ where the  matrices $\mathbf{A} = \int_{CY} \beta \wedge *_6 \, \alpha$ and $\mathbf{C} = \int_{CY} \beta \wedge *_6 \, \beta$ are integrals over the Calabi-Yau of the three-forms $\alpha, \beta$ comprising the cohomology basis dual to the three-cycles \cite{Jockers:2004yj}.  These matrices generally possess off-diagonal entries with values set by the geometrical moduli of the compactification.}

\subsection{Photini signatures at the LHC}
The mixings between the bino and hidden-sector photini give rise to interactions between the mostly-photino neutralinos $\tilde N_i$ and the fields of the MSSM. The LHC signatures of  these interactions depend on the mixing parameters and on the photini mass spectrum. Given the absence of low energy fields charged under RR $U(1)$'s, gravity mediation is the dominant source of the photini soft SUSY-breaking masses. For the remainder of this section we will assume that gravity mediation is also the dominant source of SUSY breaking for the fields of the MSSM. In this case the photini masses are of the same order as the MSSM soft masses; this is the scenario with the richest possible phenomenology. 
Another plausible scenario---in which the dominant source of the MSSM soft masses is gauge mediation, so that all the photini are much lighter than the MSSM superpartners---will be discussed in Section~\ref{totalitarian}.

Given the expected multiplicity of the photini, on statistical grounds we may expect several (or perhaps many) of them to be lighter than the Lightest Ordinary Supersymmetric Particle (LOSP). This gives rise to 
interesting LHC signatures due to LOSP decays into photini and subsequent interphotini transitions. 
 For definiteness, in the formulae below we will concentrate on the scenario wherein the LOSP is an MSSM neutralino; it is straightforward to generalize to other cases. This does not bring any qualitatively new features except for the smallest values of  bino-photini mixing, in which case the charge of the LOSP becomes particularly significant. For these small mixings, a neutral LOSP  escapes from the detector before decaying, while a charged (or colored) LOSP may stop in the detector due to electromagnetic interactions and produce a late decay signature out-of-time with collisions.
 
\begin{figure}[t]
   \centering
   \includegraphics[width=2.5in]{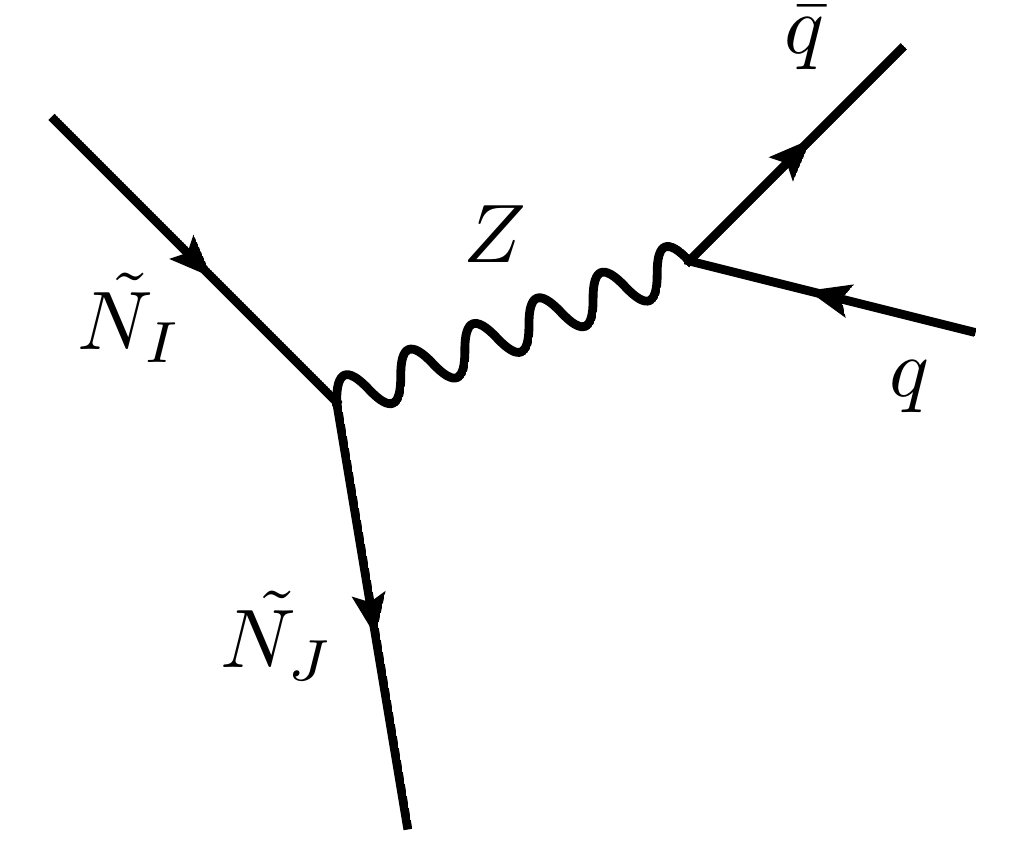} 
   \hspace{1in}
      \includegraphics[width=2.5in]{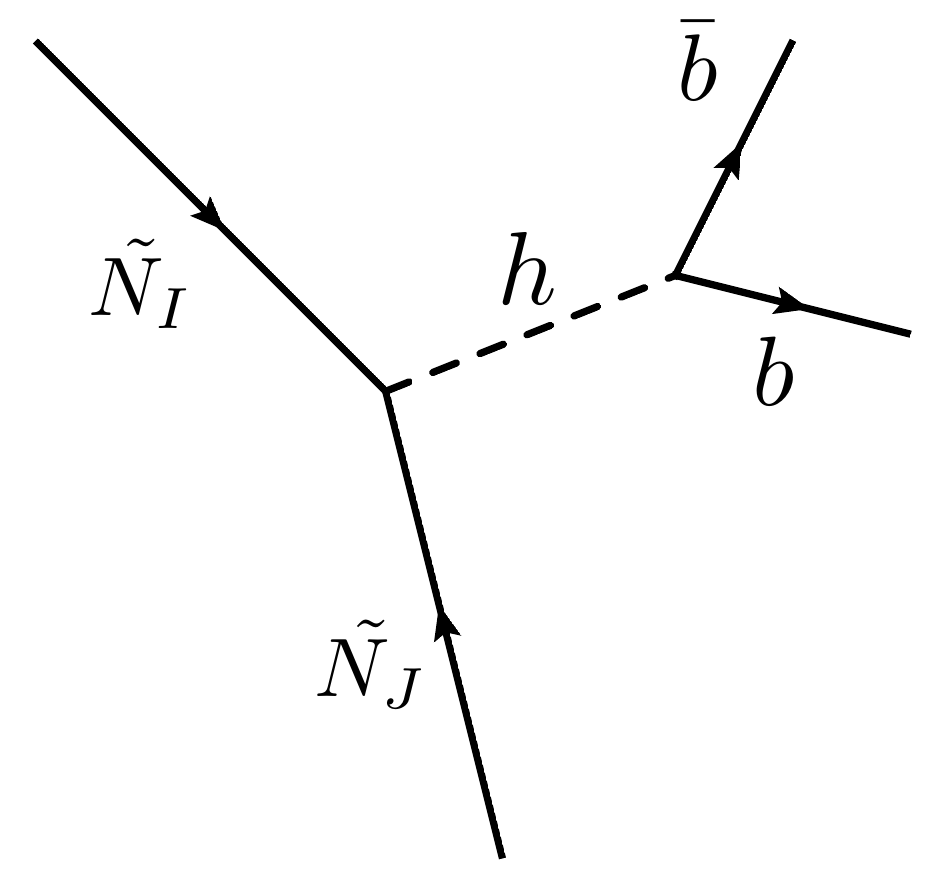}
      
      \includegraphics[width=2.5in]{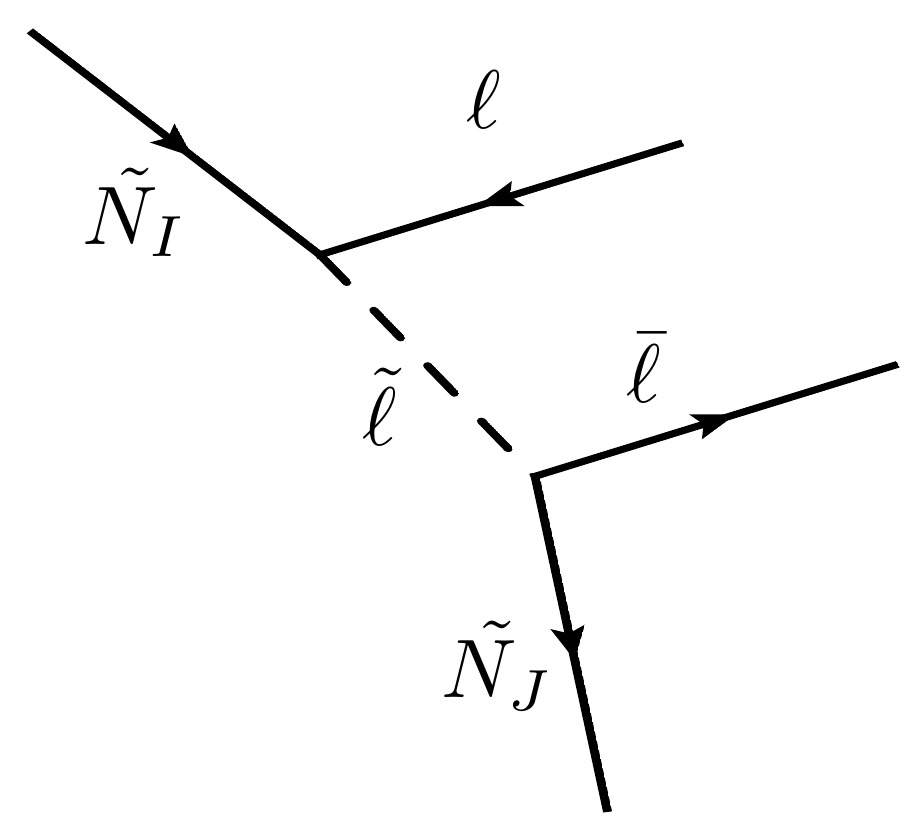}

   \caption{Different decay channels for both the LOSP decay into photini and interphotini transitions: via $Z$, Higgs, and sfermion.}
   \label{fig:inter}
\end{figure}
 
When the LOSP is an MSSM neutralino, photini production and subsequent inter-photini decays
are dominated by the following three interactions: 
\begin{enumerate}
\item Transitions through the $Z$-boson via couplings of the form $\tilde N_I \tilde N_J Z$.
\item  Transitions through the neutral Higgs $h$ via couplings of the form $\tilde N_I \tilde N_J h$.
\item  Transitions through intermediate squarks and sleptons  via couplings of the form $\tilde N_I q \tilde q$ and $\tilde N_I l \tilde l.$
\end{enumerate}
There may also be decays occurring via Standard Model photon emission, but such processes are suppressed by an additional loop factor and subdominant for a wide range of MSSM parameters \cite{Haber:1988px}. Though suppressed relative to the interactions discussed above, processes involving photon emission may constitute another noteworthy signature at the LHC. 

The decay rate $\Gamma_{IJ}^Z(\tilde N_I \rightarrow \tilde N_J + f \bar f)$ via $Z$-boson emission is parametrically of order
\be
\Gamma_{IJ}^Z(\tilde N_I \rightarrow \tilde N_J + f \bar f) \simeq 
\frac{1}{192 \pi^2} \frac{\alpha_W}{c_W^2} \left|-f_{I3} f_{J3}^* + f_{I4} f_{J4}^*\right|^2 \frac{(\delta m)^5}{m_Z^4} \, BR(Z \rightarrow f \bar f)
\ee
where $\delta m = m_I - m_J$ is the mass splitting between neutralinos (which we have assumed to satisfy $\delta m < m_Z;$ for larger splittings, the $Z$ boson is produced on-shell and two-body phase space dominates). In general, one expects decay chains ending in mostly-photino neutralinos to begin with the production of a mostly-MSSM neutralino.
For the process $\tilde N_a \to \tilde N_j,$ this corresponds to a lifetime of order
\be
\tau_Z(\tilde N_a \rightarrow \tilde N_j) \simeq 10^{-13} \text{ s } \times \left( \frac{ 10^{-2} }{\epsilon} \right)^2 \left( \frac{1}{\eta} \right)^4 \left( \frac{10 \text{ GeV} }{\delta m} \right)^5 
\ee
when $a = 1,2$ -- i.e., $\tilde N_a$ is mostly-bino or mostly-wino. Here the factor $\eta \sim \mathcal{O}(m_Z/ \mu) \sim \mathcal{O}(m_Z/m_{1,2})$ parametrizes the degree of mixing between MSSM gauginos, and may vary from $\sim 0.1-1$ depending on the size of SUSY-breaking soft masses. When $a = 3,4$ -- i.e., $\tilde N_a$ is mostly-higgsino -- the lifetime is of order $\sim \eta^2 \tau_Z(\tilde N_{1,2} \to \tilde N_j).$ The lifetime for transitions $\tilde N_i \to \tilde N_j$ between mostly-photino neutralinos is similarly given by $ \tau_Z(\tilde N_i \to \tilde N_j) \simeq \epsilon^{-2} \, \tau_Z(\tilde N_{1,2} \to \tilde N_j).$

The decay rate $\Gamma_{IJ}^{h}(\tilde N_I \rightarrow \tilde N_J + f \bar f)$ via the Higgs $h$ goes like 
\be
\Gamma_{IJ}^{h}(\tilde N_I \rightarrow \tilde N_J + f \bar f) \simeq 
 \frac{1}{192 \pi^3}  \left| - Q_{IJ} \sin \alpha - S_{IJ} \cos \alpha\right|^2 \frac{(\delta m)^5}{m_h^4} \, BR(h \rightarrow f \bar f)
\ee
where $Q_{IJ} = \frac{1}{2} [f_{I3} (f_{J2} - t_W f_{J1}) + f_{J3} (f_{I2} - t_W f_{I1})],$ $S_{IJ} = \frac{1}{2} [f_{I4} (f_{J2} - t_W f_{J1}) + f_{J4} (f_{I2} - t_W f_{I1})],$ and $\alpha$ is the usual angle of rotation between the neutral Higgs mass eigenstates. 

This corresponds to a lifetime for $\tilde N_a \rightarrow \tilde N_j$ of order
\be
\tau_{h}(\tilde N_a \to \tilde N_j) \simeq 10^{-12} \text{ s } \times \left( \frac{ 10^{-2} }{\epsilon} \right)^2 \left( \frac{1}{\eta} \right)^2 \left( \frac{10 \text{ GeV} }{\delta m} \right)^5 \left( \frac{m_{h}}{150 \text{ GeV}} \right)^4
\ee
for $a = 1,2,$ with $\eta$ as above. As before $\tau_h (\tilde N_{3,4} \to \tilde N_j) \sim \eta^2 \tau (\tilde N_{1,2} \to \tilde N_j),$ and $\tau_h(\tilde N_i \to \tilde N_j) \sim \epsilon^{-2} \, \tau (\tilde N_{1,2} \to \tilde N_j).$

The decay rate via a sfermion goes like 
\be
\Gamma_{IJ}^{\tilde l}(\tilde N_I \rightarrow \tilde N_J+ f \bar f) \simeq 
 \frac{\alpha_W^2}{48 \pi}  \left| f_{I2}^* + t_W f_{I1}^* - \frac{m_{l}}{m_W c_\beta} f_{I3}^* \right|^2 \left| f_{J2}^* + t_W f_{J1}^* - \frac{m_l}{m_W c_\beta} f_{J3}^* \right|^2 \frac{(\delta m)^5}{m_{\tilde f}^4}
\ee
which corresponds to a lifetime for $\tilde N_a \rightarrow \tilde N_j$ of order
\be
\tau_{\tilde l}(\tilde N_a \to \tilde N_j) \simeq 10^{-12} \text{ s } \times \left( \frac{ 10^{-2} }{\epsilon} \right)^2 \left( \frac{10 \text{ GeV} }{\delta m} \right)^5 \left( \frac{m_{\tilde l}}{150 \text{ GeV}} \right)^4
\ee
for $a = 1,2.$ In the case $a = 3,4,$ we have instead  $\tau_{\tilde l} (\tilde N_{3,4} \to \tilde N_j) \sim \eta^{-2}  \tau (\tilde N_{1,2} \to \tilde N_j)$ (unless $\tan \beta$ is large, in which case $\tau_{\tilde l} (\tilde N_{3,4} \to \tilde N_j) \sim  \tau (\tilde N_{1,2} \to \tilde N_j)$).  Transitions between mostly-photino neutralinos are again simply $\tau_{\tilde l}(\tilde N_i \to \tilde N_j) \sim \epsilon^{-2} \, \tau (\tilde N_{1,2} \to \tilde N_j).$ The dominant decay via sfermion exchange depends sensitively on sfermion spectroscopy; in general one expects sleptons to be lighter than squarks, thereby predominantly producing leptonic final states.

All three production mechanisms lead to parametrically similar rates. Although the decay rate via sfermion exchange is reduced at larger sfermion masses, at the same time two other channels are also being suppressed by the MSSM neutralino mixing parameter $\eta$, which is smaller for the heavier MSSM spectrum.

\begin{figure}[t]
   \centering
   \includegraphics[width=7in]{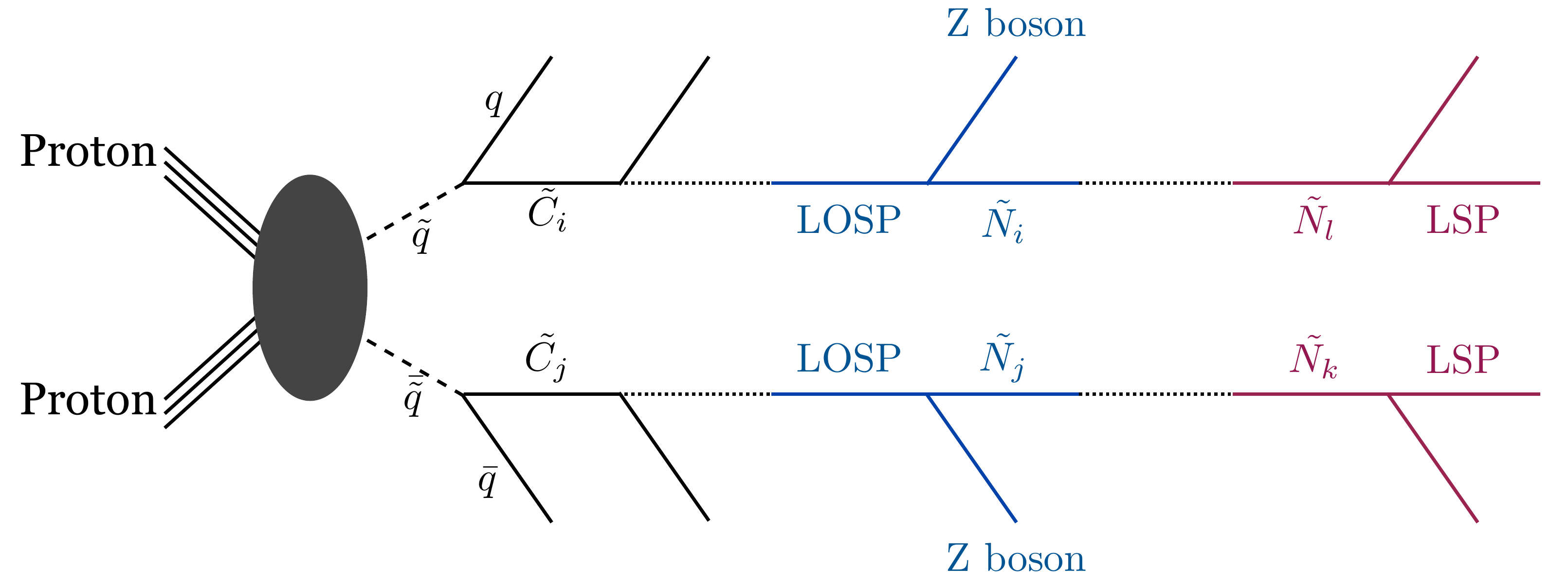} 
   \caption{The existence of multiple photini states lighter than the bino -- and mixing with MSSM neutralinos via the bino -- may modify MSSM cascade decay chains to the LSP.}
   \label{fig:cascade}
\end{figure}

\begin{figure}[t]
   \centering
   \includegraphics[width=7in]{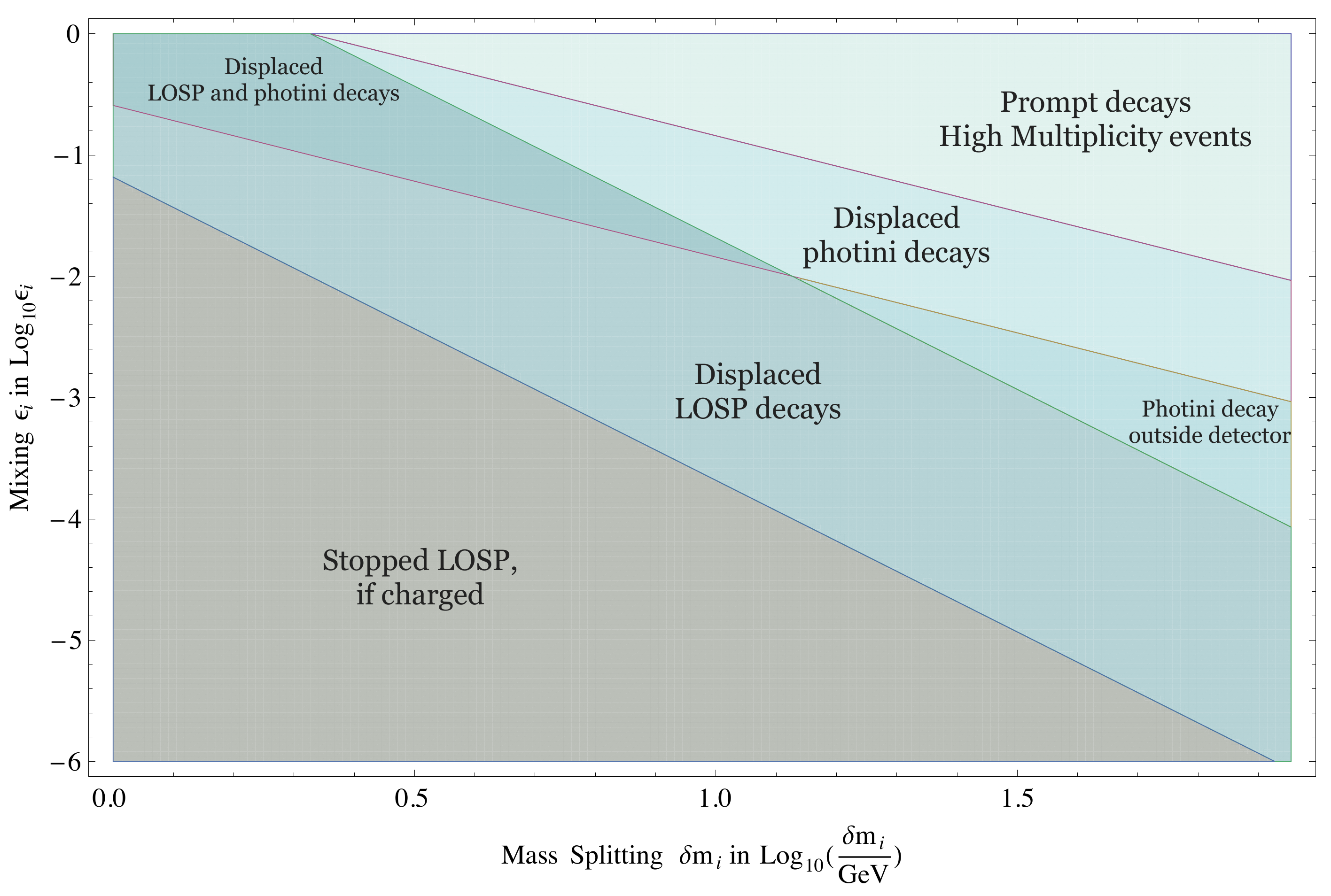} 
   \caption{The LHC signatures of multiple photini states at the LHC as a function of the photino-bino mixing $\epsilon_i$ and mass splittings $\delta m_i$.}
   \label{fig:mixplot}
\end{figure}

These processes may drastically modify MSSM particle cascades at the LHC, which no longer end at the MSSM LOSP (see Fig. \ref{fig:cascade}). Depending on the values of the mixing parameters $\epsilon_i$ and mass splittings $\delta m,$ photini give rise to several (potentially coexisting) classes of signatures illustrated in Fig.~\ref{fig:mixplot}. 

First, the usual supersymmetric cascades of the MSSM may both become longer and give rise to larger lepton multiplicities. These effects arise due to both decays of LOSP to photini and transitions among photini, either of which may happen promptly for large enough mixings and mass splittings. The branching ratios of Standard Model states produced during these transitions depend on which of the above-mentioned three decay channels dominates. In particular, decays via the $Z$ or (especially) sleptons will increase the lepton multiplicity of these cascades.  

Second, for smaller values of the mixing parameters and mass splittings one may see displaced vertices either from interphotini transitions, LOSP decays, or potentially both.

Furthermore, if interphotini transitions are too slow to be observed inside the detector it may be the case that multiple photini are discovered during  the process of mass reconstruction. In particular, if the rates of LOSP decay into several different photini are competitive, but the rate of interphotini transitions are sufficiently slow, cascades may end with photini of different masses escaping the detector. One
will then find that the observed kinematical distributions cannot be fitted by assuming a single value of the mass for the invisible particle at the end of the cascade.

For the smallest values of mixings, the signatures depend sensitively on the charge of the LOSP. Neutral LOSPs may exit the detector before decaying, resulting in rather pedestrian MSSM signatures. Charged LOSPs, however, may stop in the detector and decay out of time with collisions. As in the previous case, the interesting feature of these decays is that the mass of the invisible particle produced by late decays may vary from event to event, posing the same challenges for mass determination. This may be especially interesting for two body decays with otherwise straightforward kinematic edges, such as $\tilde{l}\to lN_i$.

It is worth stressing that some of these signatures may coexist, and the actual combination of signatures that will be observed depends on the details of the photini spectrum and mixings. For instance, if all mixing parameters are around  $\sim10^{-3}$ and the neutralino spectrum is somewhat dense, such that the mass difference between LOSP and the lightest photini is less then $\sim 30$~GeV ("Displaced LOSP decays" region in Fig.~\ref{fig:mixplot}), then one will observe
both displaced vertices from the LOSP decays and cascade decays ending in a multiplicity of invisible particles with different masses.

Another interesting possibility is that the mixing is relatively large, $\epsilon_i \gtrsim 0.05$, and the mass splittings between the LOSP and some of the photini are quite substantial, $\gtrsim 50$~GeV. This scenario -- corresponding to the "Prompt decays..." region in the upper right corner in Fig.~\ref{fig:mixplot} --
gives rise to longer prompt cascades, while a few of the lightest photini have smaller splittings (e.g., in the "Displaced photini decays" region in Fig.~\ref{fig:mixplot}) and produce displaced vertices.
 
Displaced vertices from LOSP decays or interphotini transitions are likely to provide the most striking and immediate indication of multiple photini. However, to check the distinctive feature of the axiverse---photini multiplicity reflecting the topological complexity of the underlying compactification---requires accurate photini mass determination. Furthermore, in some cases the mass determination of
the invisible particle(s) becomes the only way to distinguish this scenario from the MSSM at the LHC. Such is the case for, e.g., the "Photini decay outside detector" region in Fig.~\ref{fig:mixplot} corresponding to $\sim 10^{-3}$ mixing, in which the interphotini decays cannot be observed but prompt LOSP decays are assured by the significant splitting between the LOSP and photini.

The significant possibility of invisible final states with different masses motivates further development of mass determination techniques and their adaptation to cases in which the two decay chains in Fig.~\ref{fig:cascade} are not identical. Of particular interest are chains in which the masses of the two final invisible particles are different, as would be the case if transitions between the lightest photini happen outside the detector. This generalization is straightforward for some of the existing mass-determination techniques such as the polynomial (e.g., \cite{Kawagoe:2004rz,Cheng:2008mg})
) and the endpoint methods (e.g., 
\cite{Hinchliffe:1996iu}), but may require more work in other cases such as the $MT2$ method (e.g. \cite{Lester:1999tx,Barr:2003rg}) and its progeny (such as \cite{Burns:2008va}). On the positive side, mass determination is more efficient for longer cascades, which are to be expected in the multi-photini scenario. Ultimately, it appears realistic to expect mass determination techniques at the LHC to distinguish photini states down to splittings of $5\div 10$~GeV~\cite{private}.

The LHC phenomenology of the multi-photini scenario may be spectacular, with many leptons and displaced vertices at the end of the MSSM cascades, but it requires dedicated collider study to determine how effective ordinary SUSY searches and kinematic techniques may be in determining the parameters of this new sector. In particular,  care is required to distinguish the existence of a multi-photino sector from, e.g., NMSSM models with many singlinos. Only a measurement of the couplings between the different neutralino states (at the ILC, for example) will eventually reveal that photini couple through the bino, while for singlinos it is the Higgs that provides the bridge of communication to the MSSM.

\subsection{Collider Bounds on String Photini}

There appear to be surprisingly few collider bounds on the existence of light string photini. The customary LEP bound on the lightest neutralino mass comes from experimental limits on chargino masses and GUT relations between gaugino masses, the relaxation of which leaves few constraints on the mass and mixings of light neutralinos \cite{Dreiner:2009ic}. Potential bounds on the parameters of photini states can come either from precision measurements of the $Z$ width or direct production, since there are no light states with RR $U(1)$ charge.

The best current bounds on photino production come from LEP direct search limits on processes like $e^+ e^- \to \tilde N_1 \tilde N_2$. The model-independent bounds from LEP OPAL searches at $\sqrt{s} = 208$ GeV constrain $\sigma(e^+ e^- \to \tilde N_1 \tilde N_2) \times BR(\tilde N_2 \to Z \tilde N_1) \lesssim 70$ pb for $m_1 + m_2 < 208$ GeV \cite{Abbiendi:2003sc}. This amounts to a relatively weak constraint on photino masses and mixings for all but the largest values of $\epsilon;$ the bounds are negligible even for $\epsilon \sim 1$ provided sufficiently heavy sleptons and small higgsino-photino mixing. 

Precision electroweak observables may provide another probe of string photini. Among other quantities, the invisible $Z$ width is a sensitive probe of additional light states. If there are $N$ photino states that the $Z$ boson can decay to, the contribution to its decay width is given by \cite{Heinemeyer:2007bw}

\begin{eqnarray}
\delta \Gamma_Z \simeq \frac{G_F}{6 \sqrt{2} \pi} m_Z^3 \sum_{i,j=1}^N (\epsilon_i \epsilon_j)^2 (f_{i4} f_{j4}^* - f_{i3} f_{j3}^*)^2 
 \sim  \frac{G_F }{6 \sqrt{2} \pi} N^2 \epsilon^4 \eta^4 m_Z^3 \\
 \sim 0.03~\mbox{MeV} \times N^2  \left(  \frac{\epsilon}{0.1}\right)^4 \left( \frac{\eta}{1} \right)^4
\end{eqnarray}

Given that the invisible Z decay width has been measured with an error of 1.5 MeV \cite{Amsler:2008zzb}, photini states that are lighter than half the Z mass, i.e. 45 GeV, are constrained to have a combined mixing with the Higgsinos smaller than
\begin{eqnarray}
\epsilon \eta \lesssim \frac{0.3}{\sqrt{N}},
\end{eqnarray}
which is relevant only in the case where there are many photini lighter than $m_Z/2$ with $\mathcal{O}(1)$ mixing to the Standard Model. 

Although there are many other potential constraints from existing Standard Model parameters (including, e.g., corrections to the $W$ mass, $\sin^2 \theta_W,$ EDMs, muon $g-2$, and rare meson decays), such constraints are no stronger than the relatively weak bounds discussed above.

\section{Cosmology of String Photini}
\label{cosmo}

The cosmological implications of multiple photini coupled to the MSSM through the hypercharge portal may be problematic (cf. \cite{Ibarra:2008kn}). Even if inflation does not reheat these states directly, they will be thermalized by MSSM interactions provided $\epsilon_i \gtrsim 10^{-6}$. If a photino is the LSP, it will generically exceed the observed dark matter relic abundance by a prohibitive amount. 

A photino LSP $\tilde \gamma$ may freeze-out while nonrelativistic for mixings of $\epsilon \gtrsim 10^{-3}.$ However, in this case the photino will be overabundant by a factor of $\sim \epsilon^{-4}$ (an unfortunate consequence of the convenient fact that weak interactions alone may produce the observed dark matter relic abundance). On the other hand, for $\epsilon < 10^{-3}$ their interactions will freeze-out while the photini are relativistic, so that the photino LSP will dominate over SM radiation at $T\sim m_{\tilde \gamma}$ for $10^{-6} <\epsilon <10^{-3}$. When $\epsilon \lesssim 10^{-6}$, the photini do not reach thermal equilibrium with the MSSM, but out-of-equilibrium photino production will nonetheless overclose the universe with photini by an amount $\propto \left( \frac{\epsilon}{10^{-11}}\right)^2$. Clearly, some mechanism is necessary to dilute the photini overabundance for a vast range of mixings.\footnote{It is worth noting that the massless photons of these hidden $U(1)$s remain cosmologically irrelevant (provided they are not direct products of the inflaton's decay), since there are no light states charged under them and gravitational interactions alone will not lead to their overproduction.} Of course, these constraints are far from ironclad. In what follows we will see how the challenges of photino cosmology may be overcome in a variety of ways.

\subsection{Photino {\it qua} LSP}

As observed above, the freeze-out of a nonrelativistic photino LSP generally leads to an overabundance of order $\epsilon^{-4}.$ However, it is nonetheless possible to obtain a suitable photino relic abundance from conventional freeze-out in a proscribed region of parameter space. For sufficiently large values of $\epsilon$, coannihilation with MSSM higgsinos may lead to a freeze-out relic abundance compatible with observations. 

Higgsino dark matter is well known to yield low relic abundance due to its efficient annihilation into gauge bosons and coannihilation with charginos. If the photino LSP is sufficiently close in mass to the higgsino (i.e., provided $(m_{\tilde H} - m_{\tilde \gamma})/m_{\tilde \gamma} \lesssim T_f / m_{\tilde \gamma} \sim 5 \%$) it may coannihilate efficiently at freeze-out with an appreciable abundance of higgsinos. The coannihilation cross section scales as $\epsilon^2,$ and the resultant photino relic abundance is approximately $\Omega_{\tilde \gamma} h^2 \simeq 0.1 \,  \left( \frac{10^{-1}}{ \epsilon} \right)^2 \, \left(\frac{\mu}{100 \text{ GeV}} \right)^2.$  A similar scenario may arise by coannihilation with the stau, again provided a correlation between masses within $\sim 5 \%.$ 

Of course, the overproduction of photino dark matter for $\epsilon \lesssim 0.1$ may be ameliorated if the photini are themselves never in thermal equilibrium with the MSSM. Scattering processes that maintain photini in thermal equilibrium become inefficient below $\epsilon \lesssim 10^{-6}.$ However, even if they are not in thermal equilibrium, an appreciable abundance of photini may still be generated via interactions of MSSM particles in the thermal bath \cite{FIMP}. The resulting relic abundance from thermal production is relatively insensitive to the reheating temperature (as the photini couple to the MSSM via renormalizable interactions), and scales as $\Omega_{\tilde \gamma} h^2 \simeq 0.1 \, \left(\frac{m_{\tilde \gamma}}{100 \text{ GeV}} \right) \, \left(\frac{\epsilon}{10^{-11}} \right)^2.$ Even photini that do not reach thermal equilibrium will be prohibitively overproduced by thermal production for all but the smallest mixings.

However, if the reheating temperature $T_R$ following a period of entropy production is below the photino freeze-out temperature $T_f,$ then the relic abundance may be significantly reduced \cite{Giudice:2000ex}.  While it is possible for the primary period of inflation to end with such low $T_R,$ it would be difficult to account for baryogenesis or the observed cosmological density perturbations. A more palatable cosmological history might involve a second phase of weak-scale \cite{Randall:1994fr} or thermal \cite{Lyth:1995ka} inflation at lower energies. Such cosmologies reconcile a low $T_R$ with baryogenesis and density perturbations, and may be further required to resolve any additional moduli problems. The upper bound on $T_R$ required to avoid excess thermal production is imposed by the requirement that MSSM superpartners not reach thermal equilibrium after reheating.

Of course, another possibility is simply that an MSSM neutralino is the LSP. Such a scenario is not unreasonable if the gauginos all obtain SUSY-breaking masses from a single source, since RG running may lower the masses of MSSM gauginos relative to those of the photini. In this case, all the conventional considerations for MSSM neutralino relic abundance still pertain. Such a scenario leads to unpromising photino signatures at the LHC for all but the largest values of $\epsilon$; only for $\epsilon \gtrsim 0.1$ and $m_{\tilde \gamma} \sim m_{LSP}$ would the hidden-sector photini be expected to appear in sparticle cascades if a photino is not the LSP.

\subsection{Photino decay into a nonthermal sector}

The photino overabundance problem may also be ameliorated if the photini decay to a lighter R-parity odd state that was never in thermal equilibrium and does not dominate the energy density of the universe. In order for this to occur, it is necessary both for the photini to decay before their energy density dominates over radiation, and for the mass $m_{LSP}$ of the R-parity odd particle to be sufficiently small. For $\epsilon \gtrsim 10^{-3}$ and $10^{-6} \gtrsim \epsilon \gtrsim 10^{-11}$ these requirements suggest that the lightest photino decay rate is $\Gamma > H(T_{eq}/\epsilon)$ in the former case and $\Gamma > H(T_{eq} \epsilon^{1/2}/10^{-11/2})$ in the latter case, where $T_{eq}\sim 1$ eV is the temperature at matter-radiation equality and $\frac{m_{LSP}}{m_{\tilde{\gamma}}}< \frac{1}{\epsilon^4}$ and $\frac{m_{LSP}}{m_{\tilde{\gamma}}}< \frac{\epsilon^2}{10^{-22}}$, respectively. If the decay involves any MSSM particles, the lifetime must not exceed one second in order to preserve successful BBN predictions. Finally, for $10^{-3} \gtrsim \epsilon \gtrsim 10^{-6}$ the lightest photino decouples while relativistic, so that we require $\Gamma < H(m_{\tilde{\gamma}})$ and $m_{LSP} < 0.1$ eV. 

A promising candidate for such an R-parity odd particle may be an axino $\tilde a$ that couples to photini through interactions of the form $\frac{\alpha}{4 \pi f_a}\tilde{a} \tilde{\gamma}_i \sigma_{\mu \nu} F_i ^{\mu \nu}$. The mixing of the photini to the bino implies a decay channel $\tilde{\gamma}_i \rightarrow \tilde{a} + \gamma$, so that the lifetime has to be faster than 1 sec -- i.e., $\frac{\alpha}{4 \pi f_a}< 10^{-13} \text{ GeV}^{-1}$.  The axino mass is naturally $\sim m_{3/2};$ lighter masses require a no-scale SUSY-breaking scenario that itself may be spoiled by radiative corrections. Even in the no-scale case, the very coupling that induces photino decay generates an irreducible one-loop contribution to the axino mass of order $m_{\tilde{a}}\sim \frac{m_{\tilde{\gamma_i}}}{16 \pi^2} \big( \frac{\alpha}{4 \pi f_a}\big)^2 \Lambda^2\sim 10^{-8}m_{\tilde{\gamma_i}} \big(\frac{\Lambda}{f_a}\big)^2$, where $\Lambda$ is the smaller of the SUSY and the PQ breaking scale. As a result, the axino is unlikely to solve the photini overabundance problem for $\epsilon< 10^{-2}$.

\subsection{Photino decay into a thermal sector}

Finally, the photini overabundance problem may be solved if photini can decay before BBN
into a sector (hidden or visible) that is in thermal equilibrium with the primordial plasma at the time of the decay. A hidden sector of this genre may naturally arise from a distant stack of D-branes on the compactification manifold. To make the photino decay possible it should contain, e.g., a pair of R-parity even chiral superfields $h,\bar{h}$ charged under the hidden sector $U(1)$ group.
Then the photino decay will proceed through the mixing of the RR photini with the hidden sector neutralino.
The $\mu$-term that determines the mass of the fermionic components $\psi_{h,\bar h}$ must be in the range $\sim 1$~MeV$\div 1$~GeV, in which case the lightest photino may decay into a scalar-fermion pair $\chi_h\to h\psi_{\bar h}$ through the gauge interactions. Since the scalar $h$ is likely to acquire a significant soft mass from SUSY breaking, we also require a superpotential Yukawa interaction allowing the scalar to decay into a pair of hidden fermions. A toy example of such a hidden sector with the required properties would be a mirror sector with the MSSM field content but  a somewhat smaller $\mu$-term. The precise decay time depends on the mixing parameters, hidden sector Yukawas and scalar masses, but naturally happens before BBN since it proceeds through renormalizable interactions.

If the hidden photon is massless (or light, with a mass of order the hidden fermion masses) and mixes with the SM photon with the mixing parameter $\gtrsim 10^{-8},$ the hidden sector will remain in thermal equilibrium with the MSSM until the hidden fermions freeze out. The lower bound on the hidden fermion mass comes from the requirement that this freeze-out occurs before  BBN. This scenario is just a supersymmetric version of the usual paraphoton scenario with the hidden fermions as light millicharged particles; existing bounds on millicharged particles leave a large region in the mass/mixing parameter space for this scenario to work.

The presence of the extra hidden sector may or may not have a significant impact on LHC phenomenology. An interesting scenario may arise if the rate for photino decay into the hidden sector is faster than the rate for transition between photini, while the MSSM neutralino decays preferentially into the RR photini. In this case,  the displaced vertices due to the interphotini transitions are absent, but multiple missing final states with different masses still serve as a signature of the axiverse.

An equally viable scenario would involve the MSSM itself as the thermal sector, where the decay proceeds into Standard Model fermions through R-parity breaking operators. In lieu of exact R-parity, another anomaly-free discrete symmetry such as baryon triality could forbid dimension-four and -five operators leading to the proton decay while allowing the lepton-violating interactions $LLE, QLD$. The strongest bounds on some of the R-parity violating Yukawas in this case come from the neutrino masses at the level $3\times 10^{-6}$ (while some of the new Yukawas are practically unconstrained). Again, depending on the values of the new Yukawas and mixing parameters, these new interactions may either eliminate the LHC signatures of photini (e.g., if the new Yukawas are large and the would-be MSSM LSP  decays immediately) or leave them completely unchanged (e.g., if all new Yukawas are at the level $10^{-5} \div 10^{-6}$, and the mixing between photini is at the level $\epsilon \sim 10^{-2}$).

\section{Light Photini from Gauge Mediation} 
\label{totalitarian}

Thus far we have focused largely on theories where both MSSM fields and RR photini gain weak-scale soft masses from conventional gravity mediation. However, it is entirely possible that the primary communication of SUSY breaking to the Standard Model occurs through gauge interactions. Since the messengers of gauge mediation are assuredly not charged under the RR $U(1)$ gauge groups, gravitational effects are the sole source of photino masses and mixings; the natural value of photino masses is then $m_i \sim m_{3/2}.$ Preserving the successful flavor-blindness of gauge mediation suggests that gravity-mediated contributions to soft scalar masses-squared are no more than one part in one thousand. On the other hand, sparticle mass limits require the messenger masses to exceed $\sim 10$ TeV. Taken together, this implies that the gravitino and photino masses in a gauge-mediated scenario may be expected to range from $m_{3/2}, m_i \sim 0.1 \text{ eV} \div 1 \text{ GeV}.$

When the photino masses are particularly small, the bino-photino mixing is diminished further by the ratio of masses so that $f_{i1} \simeq \epsilon_i \frac{m_{i}}{m_B}.$ This suggests that the effective mixing given by $f_{i1}$ may be significantly smaller than the intrinsic mixing $\epsilon_i$; for a gauge-mediated scenario, the expected range of photino masses implies $10^{-12} \lesssim f_{i1} \lesssim 10^{-2}$ for $\epsilon_i = 1.$

The LHC signatures of very light photini may differ from those of their heavier brethren. Interphotini transitions are suppressed by a factor of $\left( \epsilon \frac{m_{\tilde \gamma}}{m_B} \right)^4$ and thus quite unlikely to produce observable particle cascades. The same smallness of effective mixing between light photini and the MSSM does, however, increase the likelihood of displaced vertices.  The lifetime for the decay of an MSSM neutralino LOSP to a light photino scales as 
\be
\tau(\tilde N_a \to \tilde N_i) \sim 10^{-8} \text{ s} \times \left(\frac{10^{-2}}{\epsilon} \right)^2 \left( \frac{1 \text{ MeV}}{m_{i}} \right)^2 \left( \frac{100 \text{ GeV}}{m_a} \right)^3.
\ee
Such decay of a neutral LOSP in the detector will result in displaced vertices and missing energy from photini escaping the detector. In this case, however, the mass splittings among photini are far to small to be resolved with available mass resolution, so that all indications of multiplicity are lost. The lifetime may also be sufficiently long for the neutral LOSP to escape from the detector entirely before decaying, resulting in no deviations from the conventional MSSM phenomenology. 

When SUSY breaking is communicated by gauge mediation, it is quite likely that the MSSM LOSP is charged -- as occurs frequently with the stau for lower values of the messenger scale. The case of a charged LOSP remains exceptionally interesting for even the longest of lifetimes, as the LOSP is likely to  stop in the detector due to electromagnetic interactions before decaying out of time into photini.

Naturally, the LHC signatures of light photini bear a superficial resemblance to those of a conventional gravitino LSP in theories of gauge mediation. Indeed, the rates for decays into photini and the gravitino may be competitive. The relative rates scale as
\bea
\frac{\Gamma(\tilde N_a \to \tilde N_a + ...)}{\Gamma(\tilde N_a \to \tilde G + ...)} \sim \frac{1}{12 \pi} \frac{\alpha_W}{c_W^2} \epsilon_i^2 \frac{m_i^2 m_{3/2}^2 M_P^2}{m_B^2 m_Z^4} 
\eea
where we have assumed a mostly-bino neutralino LOSP. This suggests that neutral LOSP decay to photini dominates over decays to the gravitino when $\epsilon_i \frac{m_i^2}{m_W^2} \gtrsim 10^{-15}$ (e.g., for $\epsilon_i \gtrsim 10^{-5}$ when $m_i \sim$ MeV -- a wide range of parameters). 

Discriminating between the two cases is largely a matter of branching ratios. For example, photon production via $\tilde N_a \to \gamma + \tilde G$ is the dominant channel for the decay of a bino-like neutralino into a gravitino, while decays involving $Z$ or Higgs are suppressed by factors of $(1 - m_Z^2/m_a^2)^4$ and $(1- m_h^2/m_a^2)^4,$ respectively. In contrast, the decay of a bino-like neutralino into photini proceeds dominantly via the $Z$ or Higgs, while the decay into (Standard Model) photons $\tilde N_a \to \gamma + \tilde N_i$ is suppressed by an additional loop factor. 

The cosmology of such light photini is, as one might expect, somewhat delicate. Even if the intrinsic mixing $\epsilon$ is large, the effective mixing is bound to be significantly smaller. Consequently, if a photino is the LSP there is little hope of attaining an appropriate relic abundance from freeze-out. Indeed, if the photini ever achieve thermal equilibrium with the MSSM, they will generically freeze-out while relativistic and remain subject to the usual constraints on hot relics. A more likely scenario is that these photini never achieve thermal equilibrium (as is the case, e.g., for $\epsilon \lesssim 10^{-1}$ when $m_{\tilde \gamma} \sim$ MeV), though they may be overproduced by scattering in the thermal bath. Owing to the smallness of the effective mixing, however, the resulting relic abundance may be suitable for a far greater range in $\epsilon;$ for light photini the abundance from thermal production is approximately 
\be
\Omega_{\tilde \gamma} h^2 \simeq 1.0 \times \left( \frac{\epsilon}{10^{-3}} \right)^2 \left( \frac{m_{\tilde \gamma}}{1 \text{ MeV}} \right)^3.
\ee
In cases where the photini are overproduced, the mechanisms discussed in Sec.~\ref{cosmo} may still be effective, albeit at significantly lower scales. In any case, the longevity of the MSSM LOSP may also be a problematic source of late decays; for too small values of $\epsilon$ (e.g., $\epsilon \lesssim 10^{-6}$ for $m_{\tilde \gamma} \sim$ MeV), the LOSP decay may spoil the successful predictions of Standard Model BBN. A long-lived charged or colored LOSP would be further constrained by the CMB and heavy element searches. The decay of the gravitino itself into photini and hidden-sector photons is relatively uninteresting owing to the lightness of the gravitino, occurring with a lifetime far exceeding the age of the universe: $\tau(\tilde G \to \tilde \gamma_i + \gamma_i) \sim 10^{23}  \left( \frac{m_{\tilde G}}{1 \text{ MeV}} \right)^3 \text{ s}.$ In this case, the longevity of the gravitino implies that the usual gauge mediation constraints on gravitino cosmology must be respected, even though the gravitino is not the LSP.

If, instead, the gravitino is the true LSP, very little changes; decays of the MSSM LOSP still occur preferentially into photini and Standard Model fields for a wide range of parameters. The decay of the lightest photino into a gravitino and hidden sector photon is extremely slow, also on the order of $\tau(\tilde N_i \to \tilde G + \gamma_i)  \sim 10^{23}  \left( \frac{m_{\tilde G}}{1 \text{ MeV}} \right)^3 \text{ s}.$ Both the lightest photino and the gravitino are cosmologically long-lived, and conventional considerations regarding their abundances and impact on Standard Model cosmology still apply\footnote{Note, however, that both in this and the previous cases, some of the constraints will be modified, because gravitinos are not being produced from the LOSP decays, that lead instead to the photini production.}.

Another potential cosmological bound on light photini with masses $\lesssim 30$ MeV may come from supernova cooling via photino-strahlung. The pair-production of light photino states is highly suppressed, however, and readily satisfies constraints from SN1987a \cite{Dreiner:2003wh} due to the small effective mixing, $\epsilon \frac{m_{\tilde \gamma}}{m_B} \lesssim 10^{-5}.$ For example, the so-called ``Raffelt criterion''  -- that exotic cooling processes do not alter the observed neutrino signal provided their emissivity is sufficiently small -- requires $\dot{\mathcal{E}} \lesssim 10^{19}$ ergs/g/s. For light photini,  the emissivity of photino-strahlung via slepton exchange scales as $\dot{\mathcal{E}} \sim 10^{19} \left( \frac{100 \text{ GeV}}{m_{\tilde e}} \right)^4 \left( \frac{\epsilon \, (m_{\tilde \gamma}/m_B)}{0.1} \right)^{4}$ ergs/g/s, consistent with observations for $\epsilon \frac{m_{\tilde \gamma}}{m_B}\lesssim 10^{-1}.$

\subsection{Astrophysical signatures of light photini}

Although it is important that no decays occur around the time of BBN, transitions among light photini may be slow enough to occur on cosmologically interesting timescales. The production of Standard Model particles during interphotini transitions may be observable and, moreover, well-suited to explain observed astrophysical anomalies associated with MeV-scale physics. The 511 keV excess associated with $e^+ + e^-$ annihilations recently measured by the INTEGRAL satellite \cite{Knodlseder:2003sv, Jean:2003ci, Knodlseder:2005yq} may be explained by just such transitions. It is crucial that electrons and positrons produced by the decay of a dark matter particle not be injected with more than a few MeV of energy in order to fit existing gamma ray backgrounds \cite{Beacom:2005qv}, a constraint easily satisfied by transitions among photini with masses and splittings of order $\sim$ MeV. The photon flux measured by INTEGRAL may be accounted for by a dark matter particle of mass $m$ and abundance $\Omega$ decaying into (among other things) $e^+ + e^-$ with a lifetime \cite{Craig:2009zv}
\be
\tau_{INT} \sim 10^{26} \left( \frac{\Omega}{0.2} \right) \left( \frac{1 \text{ MeV}}{m} \right) \text{ s}.
\ee
Amusingly, the interphotini decays via, e.g., off-shell $Z$ emission occur with a lifetime
\be
\tau(\tilde N_i \to \tilde N_j + e^+ + e^-) \sim 10^{23} \left( \frac{10^{-5}}{\epsilon} \right)^4 \left( \frac{1 \text{ MeV}}{\delta m} \right)^5.
\ee
Consequently, transitions among light photini may account for the INTEGRAL signal with $\epsilon \sim 10^{-5}$ for $m_{\tilde \gamma}, \delta m \sim$ MeV, for which the abundance from thermal production is expected to be $\Omega h^2 \simeq 10^{-4}.$ This assumes, of course, that the gravitino is not the LSP or that (invisible) decay rates into a gravitino LSP are slower than interphotini transitions. 

\section{Discussion}\label{discussion}

\subsection{Origin of mixing}

For hidden $U(1)$'s realized as either perturbative heterotic string states \cite{Dienes:1996zr} or as gauge excitations of D-branes of type-II string theory \cite{Abel:2008ai} the kinetic mixing with $U(1)_Y$, if absent at tree level, arises by a process that directly generalizes the classic calculation of Holdom \cite{Holdom:1985ag}.   For example in the type-II case stretched open string states with one end on the SM brane stack and the other on the hidden D-brane lead to massive states charged under both $U(1)$'s, and a one loop open string diagram then, in general, generates kinetic mixing \cite{Abel:2008ai}.  An interesting feature of the D-brane calculation is that it can also be interpreted as a tree-level exchange of a bulk closed string state between the two stacks, and both the NS-NS two--form $B_2$ and for Dp-branes the RR p-form $C_{p-1}$ lead to mixing.  The open string description is most useful for small separations between the stacks, while for large separation the supergravity approximation to the closed string computation is more appropriate and allows the treatment of both warped compactifications and those with fluxes.  The resulting mixing is model dependent, ranging in size from ${\cal O}(1)$ in the case of tree-level mixing, to in the loop-generated case a one-loop factor down to exponentially suppressed values if the compactification is warped, or if the mediating fields are massive, e.g., due to fluxes.

From the effective field theory point of view kinetic mixing among $U(1)_Y$ and the RR $U(1)$ is also allowed. However, one may worry that there might be a subtlety arising from the absence of perturbative string states carrying RR charges.

Indeed, as already summarized the conventional mechanism giving rise to mixing between $U(1)$ gauge bosons relies on integrating out heavy bi-fundamental fields charged under both gauge groups. However, substantial mixing between hypercharge $U(1)_Y$ and RR gauge fields
cannot be generated in this way.  Indeed, as explained above, there are no states charged under RR fields with masses below the string scale.  Moreover, the only states carrying RR charges are non-perturbative D-branes states, which should be thought of as solitonic states from the viewpoint of the hypercharge $U(1)_Y$ (which itself typically arises as a conventional perturbative string state). One may expect that loops involving such solitonic states are exponentially suppressed.\footnote{Note that in a theory with gravity there must exist Reissner-Nordstrom black holes charged under both $U(1)$'s.  Once again the contribution to kinetic mixing from integrating out these bi-fundamental states is expected to be exponentially small.}

Nevertheless, the mixing between hypercharge $U(1)_Y$ and RR gauge fields can be generated directly at the level of the string-scale supergravity effective action. One example of a situation giving rise to such a mixing was discussed in \cite{Jockers:2004yj}. Namely, one considers a D7 brane in the type IIB theory that wraps a four-cycle with a non-contractible loop inside. In the presence of a non-trivial Wilson line along this loop a mixing between the D7's perturbative $U(1)$ gauge field and RR $U(1)$ may arise from the world-volume Chern-Simons action
\[
\int_{\rm{D7}}C_4\wedge F\wedge F\;.
\]
The size of this mixing is controlled by the corresponding Wilson loop; this term takes the form of kinetic mixing when self-duality conditions are imposed on the $C_4$. Perhaps an even simpler example of such kinetic mixing arising from the Chern-Simons action occurs in the case of a D5 brane wrapping a two-cycle in type IIB string theory \cite{Grimm:2008dq}. The term of interest appears in the Chern-Simons action from the pull-back of the RR form to the world-volume of the D5 brane,   
\[
\int_{\rm{D5}} \zeta \, d C_4\wedge F\;.
\]
Here $\zeta$ is a (4d) complex scalar modulus parameterizing deformations of the D5 brane. Once again, this interaction takes the form of kinetic mixing when self-duality conditions are imposed on the $C_4.$

In these examples the D-branes serve as portals giving rise to a mixing between perturbative and RR gauge sectors. The D-brane may be either directly a part of the brane configuration giving rise to the Standard Model gauge group, or belong to the hidden sector and acquire a mixing with the hypercharge $U(1)$ at one loop in the conventional way.  

Kinetic mixing between hidden and visible $U(1)$'s then begets mass mixing; the gaugino mass matrix descends from the gauge kinetic mixing matrix when supersymmetry is broken. Properly speaking, the full gauge kinetic matrix for both photini and MSSM gauginos depends on, e.g., complex structure moduli $z_k$ (in the IIB case; the same role is played by K\"{a}hler moduli in the IIA case). The mass matrix arises when the complex structure moduli are replaced by their $F$-term expectation values, so that $m_{IJ} \propto F_{z_k} \partial_{z_k} Z_{IJ}(z_l).$ In general, the $F$ terms of the various complex structure moduli are expected to vary, so that the mass matrix $\mathbf{m}$ is not strictly proportional to the kinetic mixing matrix $\mathbf{Z}.$ Likewise, the size of mass mixings may exceed the size of kinetic mixings, so that $\epsilon \sim \mathcal{O}(1)$ mixings in the gaugino mass matrix may remain consistent with perturbative gauge coupling unification.

This discussion strongly suggests, that the effective field theory expectation is correct; there is no obstruction for the mixing between RR gauge fields and hypercharge, just as there is no obstruction for mixing between two D-brane $U(1)$'s. Nevertheless, it is worth keeping in mind that,
to the best of our knowledge, there is no explicit example of string theory vacuum supporting this point. For instance, it turns out that the 
D7 mechanism above doesn't give rise to non-zero mixing for the toric Calabi-Yau's (the only explicitly studied example), due to apparently accidental cancelation \cite{Hans}. 

We don't think this is a reason to worry that the phenomenology discussed in this paper may be disfavored---it appears likely that the lack of explicit examples is just a reflection of the well-known fact, that constructing explicit 
string vacua with stabilized moduli is hard.
At the very least, as explained in the introduction, extra $U(1)$ factors with no light charged states may come also from hidden D-branes, rather than from
RR fields. Rather, we consider this theoretical problem as a motivation for further studies of the plausible sources and sizes of the mixing.

\subsection{The scale of SUSY breaking}

Throughout this paper we have focused on the observational consequences of string photini in a conventional low energy SUSY scenario. However, such photini may also be observed at the LHC in a high-scale scenario such as split supersymmetry \cite{ArkaniHamed:2004fb, Giudice:2004tc, ArkaniHamed:2004yi}. In this case, scalar superpartners are heavy and inaccessible at the LHC, but fermions remain at the weak scale due to chiral symmetries. In this scenario string photini masses are likely to remain at the weak scale due to the same R-symmetry that keeps gauginos light, rendering photini observable at the LHC. The phenomenology of split SUSY events involving the direct production of non-colored superpartners remains quite similar to the case of low energy SUSY. The main difference
between photini signatures of split supersymmetry  and conventional SUSY is the absence of  interphotini transitions through an intermediate slepton, because all sleptons are now very heavy.

A particularly interesting feature of split SUSY is that the gluino is very long lived, since all its decay channels go through a heavy intermediate squark. This results in a spectacular signature due to delayed decays of gluinos stopped in the detector. Such a signal persists in the presence of string photini, but an interesting new feature is that the wide range of photini signatures discussed earlier -- in particular, cascades and displaced vertices -- may now appear out-of-time in the decays of stopped gluinos.

\section{Conclusions}\label{conclusions}

A string-theoretic universe with small extra dimensions is often thought to leave few explicit signatures at low energies -- and particularly few signatures accessible at the LHC. Here we have seen, however, that  the topological complexity of compactification manifolds in string theory suggests the presence of many unbroken $U(1)$'s without light charged states. Despite the decoupling of the photons associated with these $U(1)$'s, contact with the Standard Model may still arise due to mixing between the photini and MSSM bino in the presence of low energy supersymmetry. This mixing gives rise to a broad range of novel signatures at the LHC, including displaced vertices and cascade decays from both LOSP decays to photini and interphotini transitions, as well as multiple reconstructed masses for particles exiting the detector. Such signatures pose new challenges to existing techniques for event reconstruction and mass determination at the LHC. Should a plenitude of photini be observed at the LHC in this fashion, it would provide compelling infrared evidence for a topologically rich string compactification in the ultraviolet.

\section*{Acknowledgments}
We thank A. Barr, J. Conlon, H. Jockers, S. Kachru, E. Palti, E. Silverstein, and S. Shenker for helpful conversations, as well as the Dalitz Institute for Fundamental Physics and the Department of Theoretical Physics, Oxford University for hospitality during the course of this work. JMR is partially supported by the EC network 6th Framework Programme Research and Training Network Quest for Unification (MRTN-CT-2004-503369), by the EU FP6 Marie Curie Research and Training Network UniverseNet (MPRN-CT-2006-035863), and by the STFC. 

\bibliography{photinirefs}

\begin{thebibliography}{99}

\bibitem{Polchinski:1995mt}
  J.~Polchinski,
  Phys.\ Rev.\ Lett.\  {\bf 75}, 4724 (1995)
  [arXiv:hep-th/9510017].

\bibitem{Green:1987mn}
  M.~B.~Green, J.~H.~Schwarz and E.~Witten,
{\it  Cambridge, Uk: Univ. Pr. ( 1987) 596 P. ( Cambridge Monographs On Mathematical Physics)}

\bibitem{Svrcek:2006yi}
  P.~Svrcek and E.~Witten,
  JHEP {\bf 0606}, 051 (2006)
  [arXiv:hep-th/0605206].


\bibitem{axiverse}
  A.~Arvanitaki, S.~Dimopoulos, S.~Dubovsky, N.~Kaloper and J.~March-Russell,
  [arXiv:0905.4720 [hep-th]].

\bibitem{Davidson:1993sj}
  S.~Davidson and M.~E.~Peskin,
  Phys.\ Rev.\  D {\bf 49}, 2114 (1994)
  [arXiv:hep-ph/9310288].

\bibitem{Davidson:2000hf}
  S.~Davidson, S.~Hannestad and G.~Raffelt,
  JHEP {\bf 0005}, 003 (2000)
  [arXiv:hep-ph/0001179].
 
\bibitem{Dubovsky:2003yn}
  S.~L.~Dubovsky, D.~S.~Gorbunov and G.~I.~Rubtsov,
  JETP Lett.\  {\bf 79}, 1 (2004)
  [Pisma Zh.\ Eksp.\ Teor.\ Fiz.\  {\bf 79}, 3 (2004)]
  [arXiv:hep-ph/0311189].
 
  
\bibitem{Melchiorri:2007sq}
  A.~Melchiorri, A.~Polosa and A.~Strumia,
  Phys.\ Lett.\  B {\bf 650}, 416 (2007)
  [arXiv:hep-ph/0703144].
  
\bibitem{Ibarra:2008kn}
  A.~Ibarra, A.~Ringwald and C.~Weniger,
  JCAP {\bf 0901}, 003 (2009)
  [arXiv:0809.3196 [hep-ph]].

\bibitem{Holdom:1985ag}
  B.~Holdom,
  Phys.\ Lett.\  B {\bf 166}, 196 (1986).

\bibitem{Dienes:1996zr}
  K.~R.~Dienes, C.~F.~Kolda and J.~March-Russell,
  Nucl.\ Phys.\  B {\bf 492}, 104 (1997)
  [arXiv:hep-ph/9610479].

\bibitem{Holdom:1990xp}
  B.~Holdom,
  Phys.\ Lett.\  B {\bf 259}, 329 (1991).

\bibitem{Jockers:2004yj}
  H.~Jockers and J.~Louis,
  Nucl.\ Phys.\  B {\bf 705}, 167 (2005)
  [arXiv:hep-th/0409098].

\bibitem{Haber:1988px}
  H.~E.~Haber and D.~Wyler,
  Nucl.\ Phys.\  B {\bf 323}, 267 (1989).

\bibitem{Hinchliffe:1996iu}
  I.~Hinchliffe, F.~E.~Paige, M.~D.~Shapiro, J.~Soderqvist and W.~Yao,
  Phys.\ Rev.\  D {\bf 55}, 5520 (1997)
  [arXiv:hep-ph/9610544].

\bibitem{Kawagoe:2004rz}
  K.~Kawagoe, M.~M.~Nojiri and G.~Polesello,
  Phys.\ Rev.\  D {\bf 71}, 035008 (2005)
  [arXiv:hep-ph/0410160].

\bibitem{Cheng:2008mg}
  H.~C.~Cheng, D.~Engelhardt, J.~F.~Gunion, Z.~Han and B.~McElrath,
  Phys.\ Rev.\ Lett.\  {\bf 100}, 252001 (2008)
  [arXiv:0802.4290 [hep-ph]].

\bibitem{Lester:1999tx}
  C.~G.~Lester and D.~J.~Summers,
  Phys.\ Lett.\  B {\bf 463}, 99 (1999)
  [arXiv:hep-ph/9906349].

\bibitem{Barr:2003rg}
  A.~Barr, C.~Lester and P.~Stephens,
  J.\ Phys.\ G {\bf 29}, 2343 (2003)
  [arXiv:hep-ph/0304226].

\bibitem{Burns:2008va}
  M.~Burns, K.~Kong, K.~T.~Matchev and M.~Park,
  JHEP {\bf 0903}, 143 (2009)
  [arXiv:0810.5576 [hep-ph]].



\bibitem{private} Alan Barr, private discussions.

\bibitem{Dreiner:2009ic}
  H.~K.~Dreiner, S.~Heinemeyer, O.~Kittel, U.~Langenfeld, A.~M.~Weber and G.~Weiglein,
  Eur.\ Phys.\ J.\  C {\bf 62}, 547 (2009)
  [arXiv:0901.3485 [hep-ph]].

\bibitem{Abbiendi:2003sc}
  G.~Abbiendi {\it et al.}  [OPAL Collaboration],
  Eur.\ Phys.\ J.\  C {\bf 35}, 1 (2004)
  [arXiv:hep-ex/0401026].

\bibitem{Heinemeyer:2007bw}
  S.~Heinemeyer, W.~Hollik, A.~M.~Weber and G.~Weiglein,
  JHEP {\bf 0804}, 039 (2008)
  [arXiv:0710.2972 [hep-ph]].

\bibitem{Amsler:2008zzb}
  C.~Amsler {\it et al.}  [Particle Data Group],
  Phys.\ Lett.\  B {\bf 667}, 1 (2008).


\bibitem{FIMP}
L.~Hall, K.~Jedamzik, J.~March-Russell, and S.~West,
[OUTP-09-18-P]. 
  

\bibitem{Giudice:2000ex}
  G.~F.~Giudice, E.~W.~Kolb and A.~Riotto,
  Phys.\ Rev.\  D {\bf 64}, 023508 (2001)
  [arXiv:hep-ph/0005123].
  

\bibitem{Randall:1994fr}
  L.~Randall and S.~D.~Thomas,
  Nucl.\ Phys.\  B {\bf 449}, 229 (1995)
  [arXiv:hep-ph/9407248].
  
\bibitem{Lyth:1995ka}
  D.~H.~Lyth and E.~D.~Stewart,
  Phys.\ Rev.\  D {\bf 53}, 1784 (1996)
  [arXiv:hep-ph/9510204].
    
\bibitem{Dreiner:2003wh}
  H.~K.~Dreiner, C.~Hanhart, U.~Langenfeld and D.~R.~Phillips,
  Phys.\ Rev.\  D {\bf 68}, 055004 (2003)
  [arXiv:hep-ph/0304289].

\bibitem{Knodlseder:2003sv}
  J.~Knodlseder {\it et al.},
  Astron.\ Astrophys.\  {\bf 411}, L457 (2003)
  [arXiv:astro-ph/0309442].
  
\bibitem{Jean:2003ci}
  P.~Jean {\it et al.},
  Astron.\ Astrophys.\  {\bf 407}, L55 (2003)
  [arXiv:astro-ph/0309484].
  
\bibitem{Knodlseder:2005yq}
  J.~Knodlseder {\it et al.},
  Astron.\ Astrophys.\  {\bf 441}, 513 (2005)
  [arXiv:astro-ph/0506026].
  
\bibitem{Beacom:2005qv}
  J.~F.~Beacom and H.~Yuksel,
  Phys.\ Rev.\ Lett.\  {\bf 97}, 071102 (2006)
  [arXiv:astro-ph/0512411].
  
\bibitem{Craig:2009zv}
  N.~J.~Craig and S.~Raby,
  arXiv:0908.1842 [hep-ph].

\bibitem{Abel:2008ai}
  S.~A.~Abel, M.~D.~Goodsell, J.~Jaeckel, V.~V.~Khoze and A.~Ringwald,
  JHEP {\bf 0807}, 124 (2008)
  [arXiv:0803.1449 [hep-ph]].

\bibitem{Grimm:2008dq}
  T.~W.~Grimm, T.~W.~Ha, A.~Klemm and D.~Klevers,
  Nucl.\ Phys.\  B {\bf 816}, 139 (2009)
  [arXiv:0811.2996 [hep-th]].

\bibitem{Hans}
Hans Jockers, private discussions.
\bibitem{ArkaniHamed:2004fb}
  N.~Arkani-Hamed and S.~Dimopoulos,
  JHEP {\bf 0506}, 073 (2005)
  [arXiv:hep-th/0405159].

\bibitem{Giudice:2004tc}
  G.~F.~Giudice and A.~Romanino,
  Nucl.\ Phys.\  B {\bf 699}, 65 (2004)
  [Erratum-ibid.\  B {\bf 706}, 65 (2005)]
  [arXiv:hep-ph/0406088].

\bibitem{ArkaniHamed:2004yi}
  N.~Arkani-Hamed, S.~Dimopoulos, G.~F.~Giudice and A.~Romanino,
  Nucl.\ Phys.\  B {\bf 709}, 3 (2005)
  [arXiv:hep-ph/0409232].


\end{thebibliography}

\end{document}